\documentclass[runningheads]{llncs}
\usepackage{amsmath}
\usepackage{graphicx}
\usepackage{url}
\usepackage[utf8]{inputenc}
\usepackage{subfig}
\usepackage{float}
\usepackage{color}
\definecolor{red}{rgb}{1.0,0.0,0.0}
\definecolor{blue}{rgb}{0.0,0.0,1}
\definecolor{green}{rgb}{0.29, 0.33, 0.13}
\begin{document}
\title{Using machine learning and information visualisation for discovering latent topics in Twitter news}
\titlerunning{A visual analytic model for news in Twitter}
%
%

%
%
\author{Vladimir Vargas-Calderón\inst{1}\orcidID{0000-0001-5476-3300} \and
 Marlon Steibeck Dominguez\inst{2} \and
 N. Parra-A.\inst{1}\orcidID{0000-0002-1829-4399}\and
 Herbert Vinck-Posada \inst{1}\and
 Jorge E. Camargo\inst{3}}
%

\authorrunning{V. Vargas-Calderón et al.}
%
\institute{Grupo de Superconductividad y Nanotecnología, Departamento de Física, Universidad Nacional de Colombia, AA 055051, Bogotá, Colombia 
\email{\{vvargasc, nparraa, hvicnkp\}@unal.edu.co}\\\and
Facultad de Ingenier\'ia, Unipanamericana Fundaci\'on Universitaria, 111321, Bogot\'a, Colombia \email{madominguez@unipanamericana.edu.co} \and Departamento de Ingeniería de Sistemas e Industrial, Universidad Nacional de Colombia, AA 055051, Bogotá, Colombia \email{jecamargom@unal.edu.co}}

\maketitle              
\begin{abstract}
We propose a method to discover latent topics and visualise large collections of tweets for easy identification and interpretation of topics, and exemplify its use with tweets from a Colombian mass media giant in the period 2014--2019. The latent topic analysis is performed in two ways: with the training of a Latent Dirichlet Allocation model, and with the combination of the FastText unsupervised model to represent tweets as vectors and the implementation of K-means clustering to group tweets into topics.  Using a classification task, we found that people respond differently according to the various news topics. The classification tasks consists on the following: given a reply to a news tweet, we train a supervised algorithm to predict the topic of the news tweet solely from the reply. Furthermore, we show how the Colombian peace treaty has had a profound impact on the Colombian society, as it is the topic in which most people engage to show their opinions.



\end{abstract}
\section{Introduction}

Social interaction on social networks has been a central research topic during the last decade. It has applications in modernisation of government’s Twitter-based networking strategies~\cite{Golbeck10,Small12,Mehmet13,Suk15,Arturo18}, e-commerce~\cite{Linda10,Mata14,Zhao16} and business performance~\cite{Culnan2010HowLU,Alexandra17}.
Of particular interest is the study of how people respond to a stimulus, forming either strong collective responses or weak, isolated responses. For example, \cite{Chen12,Nargundkar16}, studied users' influence in microblogging networks, where implementing the InfluenceRank Algorithm for identifying people with high social media influence can be of great commercial value. An interesting area within the stimulus--response field in social networks is the response of people to the news. In the case of Twitter, it has been identified since its early days as a news-spreading social network~\cite{Kwak:2010:TSN:1772690.1772751} where mass media are central actors. Previous work in this area is typically focused on topics such as sports, politics, health, education, travel, business, and so on~\cite{Lee11,Zhao11,Nasser18,Garrett19}.


In this work, we inquire on the recognition of the semantic structure of people's response to news tweets by one of the largest Colombian news mass media, Radio Cadena Nacional (RCN), with (7,7 million followers by August 2019). If there is a distinction between how people respond to different kinds of news, then state of the art natural language processing and machine learning techniques should be able to detect this difference. Therefore, we propose a two-stage pipeline to investigate the claimed differences. During the first stage, topic analysis is performed over the tweets by RCN, showing the main latent topics that they tweet about. In the second stage, we take the comments by regular users to those tweets and ask ourselves if it is possible to predict which is the topic of the tweet they are responding to. Our research is two-folded. On the one hand, we investigate the characterisation of how people respond to news from different topics. On the other hand, we pay particular attention to one of the most critical social phenomena of the past decade worldwide: the response by Colombian population to the peace treaty between the Colombian government and the Colombian Revolutionary Armed Forces guerrilla (FARC) \cite{tellez2019}. These two entities were in a war for over 50 years, a conflict responsible for hundreds of thousands of deaths and millions of people forced from their homes, and being one of the largest human tragedies of modern times~\cite{cely2014grupo}.

This paper is divided as follows: in Section \ref{method} we describe in detail the dataset that we used to carry out our study as well as the workflow of the proposed method. In Section \ref{result}, we present the results and analysis of the research. Finally, the main conclusions and future work are presented in Section \ref{conclusions}.

\section{Method and Materials}\label{method}

\begin{figure}[!ht]
    \centering
    \includegraphics[width=\textwidth]{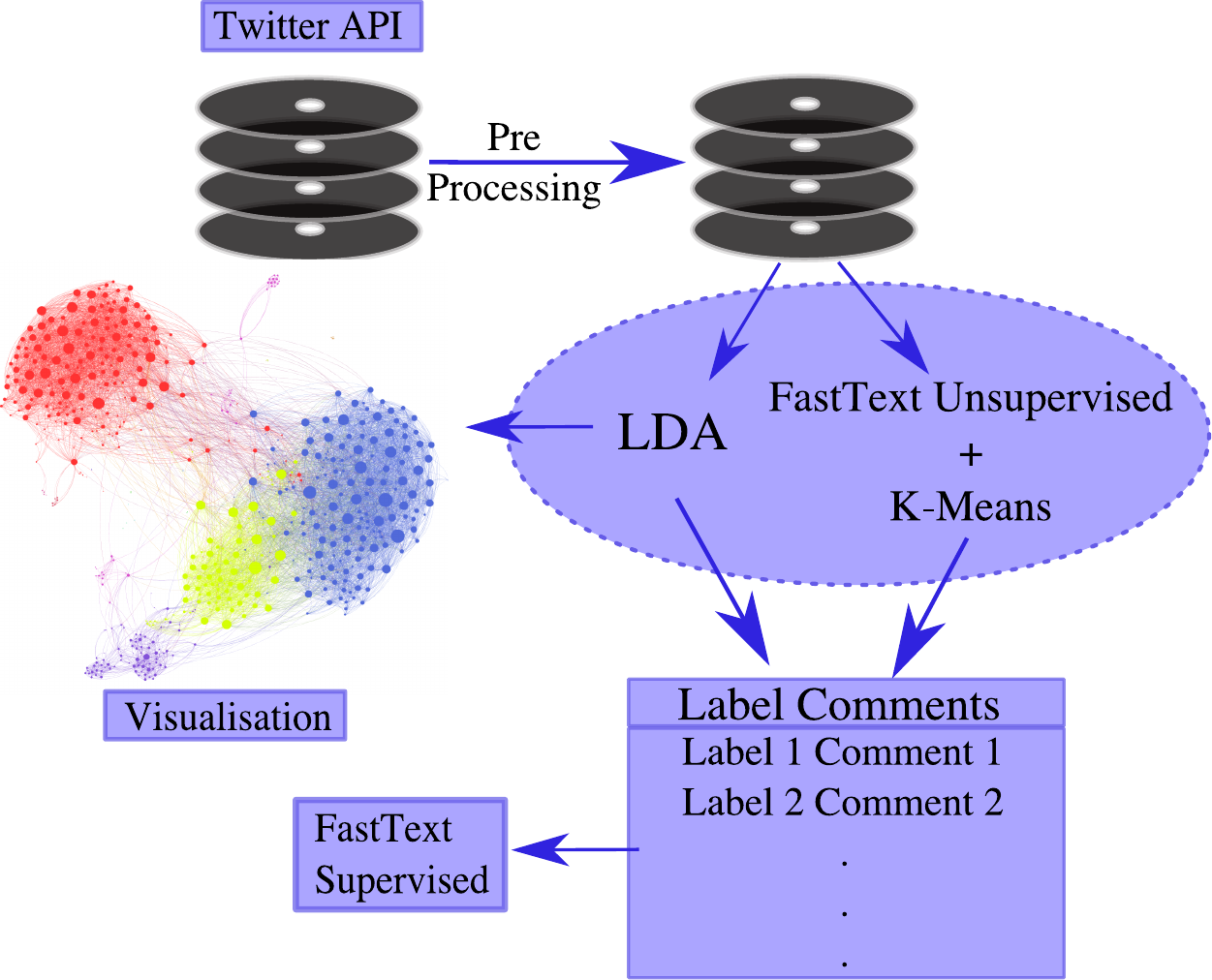}
    \caption{Overview of the proposed method. We crawl tweets from @NoticiasRCN Twitter account. Then, pre-processing is made over this set of tweets. After that, two different unsupervised topic models are trained, from which visualisations of the tweets only from @NoticiasRCN can be built. We use the discovered latent topics to label comments to @NoticiasRCN tweets. Finally, we train a supervised FastText model to predict the topics of the comments.}
    \label{fig:pipeline}
\end{figure}

In this section, we describe in detail the two stages of our research: a topic modelling and a topic prediction stage. The work pipeline is shown in Fig \ref{fig:pipeline}. 

We collected tweets from the Twitter account @NoticiasRCN from 2014 to the present using the Twitter API. During this period, the mass media giant published 258,848 tweets, having 1,447,440 comments from their public. Then, we proceeded to pre-process our tweet database. We removed punctuation, links, hashtags and mentions from all the tweets. We further lemmatised the text in lowercase.

With the pre-processed data, we trained topic modelling algorithms in order to discover topics in the tweet corpus. Particularly we used the Latent Dirichlet Allocation (LDA) model~\cite{Hoffman:2010} and a combination of unsupervised FastText model~\cite{bojanowski2016enriching,joulin2016bag} and K-Means clustering~\cite{lloyd1982least}.

\begin{figure}[!ht]
    \centering
    \includegraphics[width=\textwidth]{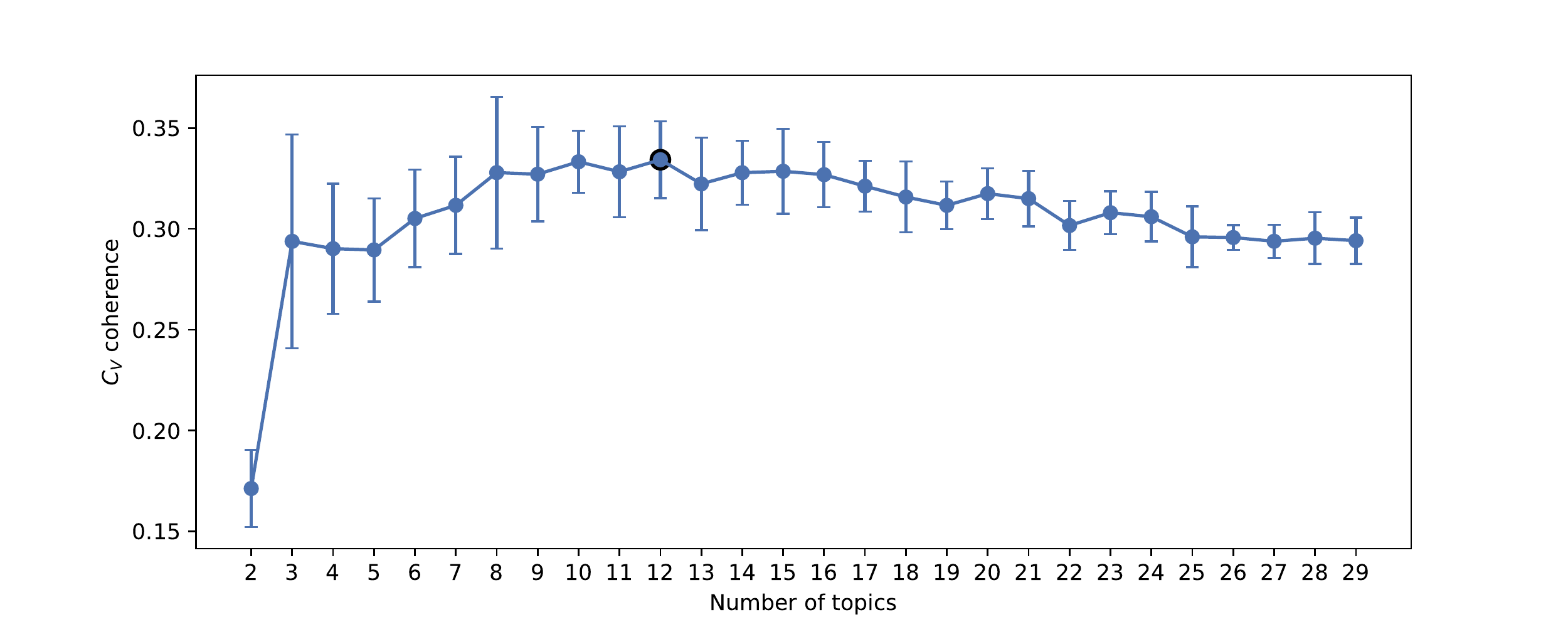}
    \caption{$C_V$ coherence as a function of number of topics for an LDA model trained with tweets from @NoticiasRCN. For each number of topics, a total of 20 LDA training runs were performed, and bars indicate a standard deviation of the corresponding 20 measured $C_V$ coherences.}
    \label{fig:cv}
\end{figure}

LDA assigns $K$ probabilities to a tweet of belonging to $K$ different classes. The classes are learnt in an unsupervised fashion from the co-occurrence of words within the tweets of the corpus. Each of the classes is called a latent topic. Therefore, the $K$ probabilities assigned to a tweet by LDA can be interpreted as a vector of $K$ components, where the $k$-th one shows the content percentage of the $k$-th topic in a tweet. The number of latent topics $K$ is a parameter that one provides to the model. In principle, each latent topic corresponds to a topic a human would understand. However, if a low number of latent topics is provided, each latent topic may contain several real topics. On the other hand, if a high number of latent topics is provided, many latent topics may refer to the same real topic. To keep a good correspondence between latent topics and human-understandable topics or real topics, the number of latent topics should be carefully selected. Refs.~\cite{Roder:2015,syed2017} have shown that the best way to pick the number of latent topics is by measuring the $C_V$ coherence, which has shown a large correlation with human judgements of the interpretability of the topics extracted by LDA. For the @NoticiasRCN tweets, the optimum number of latent topics is 12, as it is shown in Fig~\ref{fig:cv}.

The other topic modelling algorithm consisted of training an unsupervised FastText model and performing K-means clustering on the unsupervised learnt vector representations for the tweets in the corpus. FastText is a memory-efficient and fast vector embedding algorithm based on the ideas of Word2Vec~\cite{mikolov2013distributed}. An important remark is that FastText uses sub-word information to enrich word vectors. This is particularly useful in a language such as Spanish, because it is largely inflected. What vector embedding algorithms do is to represent text as vectors of $N$ components. Again, based on the co-occurrence of words or sequences of characters, FastText assigns $N$-dimensional vectors to each tweet. Here, the components do not have a human interpretation, as they embody abstract semantic features of the tweets. Once the embedded vectors for each tweet are learnt from FastText, the K-means clustering algorithm is used to group vectors in the embedded vector space. This combination of a vector embedding algorithm and a clustering algorithm has been used before with excellent results in unsupervised community detection models~\cite{vargas2019characterization}. In order to compare both topic modelling algorithms (LDA and FastText + K-means), we used K-means to form 12 clusters, where each one represents a latent topic.

After that, we labelled the comments to the news tweets. Let $t_{\text{N}}^{(i)}$ be a news tweet from @NoticiasRCN indexed by $i$, which runs from $1$ to $M$, the total number of news tweets. Let $t_{\text{C}}^{(j_i)}$ be a comment made by some Twitter user to the news tweet $t_{\text{N}}^{(i)}$ indexed by $j_i$, which runs from $1_i$ to $N^{(i)}_i$, where $N^{(i)}$ is the total number of comments to the $i$-th news tweet $t_{\text{N}}^{(i)}$ . Then, we created a labelled comment dataset for each of the two considered topic modelling algorithms as follows,
\begin{align}
    \left\{\left(\ell_1, t_{\text{C}}^{(1_1)}\right), \left(\ell_1, t_{\text{C}}^{(2_1)}\right), \ldots, \left(\ell_1, t_{\text{C}}^{(N^{(1)}_1)}\right) , \ldots, \left(\ell_M, t_{\text{C}}^{(N^{(M)}_M)}\right)\right\},
\end{align}
where $\ell_i$ is the latent topic (from either LDA or FastText + K-means) assigned to the $i$-th news tweet $t_{\text{N}}^{(i)}$. Recall that $\ell_i$ is one of the 12 different latent topics.

From the vector representation of the tweets built with the unsupervised FastText model, we generated visualisations of the tweets in a 2D map. To do this, we applied a state of the art dimensionality reduction model called the Uniform Manifold Approximation and Projection (UMAP)~\cite{2018arXivUMAP}. UMAP learns topological relations between the FastText vectors of $N$ components and finds representative projections of the data onto a two-dimensional vector space.

Finally, a supervised FastText model is trained to learn to predict the labels $\ell_i$ (provided by the two topic modelling algorithms) from the vector representations of the tweets $t_{\text{C}}^{(j_i)}$, which are efficiently learnt by FastText.

\section{Results and Discussion}\label{result}

\begin{figure}[!ht]
    \centering
    \includegraphics[width=\textwidth]{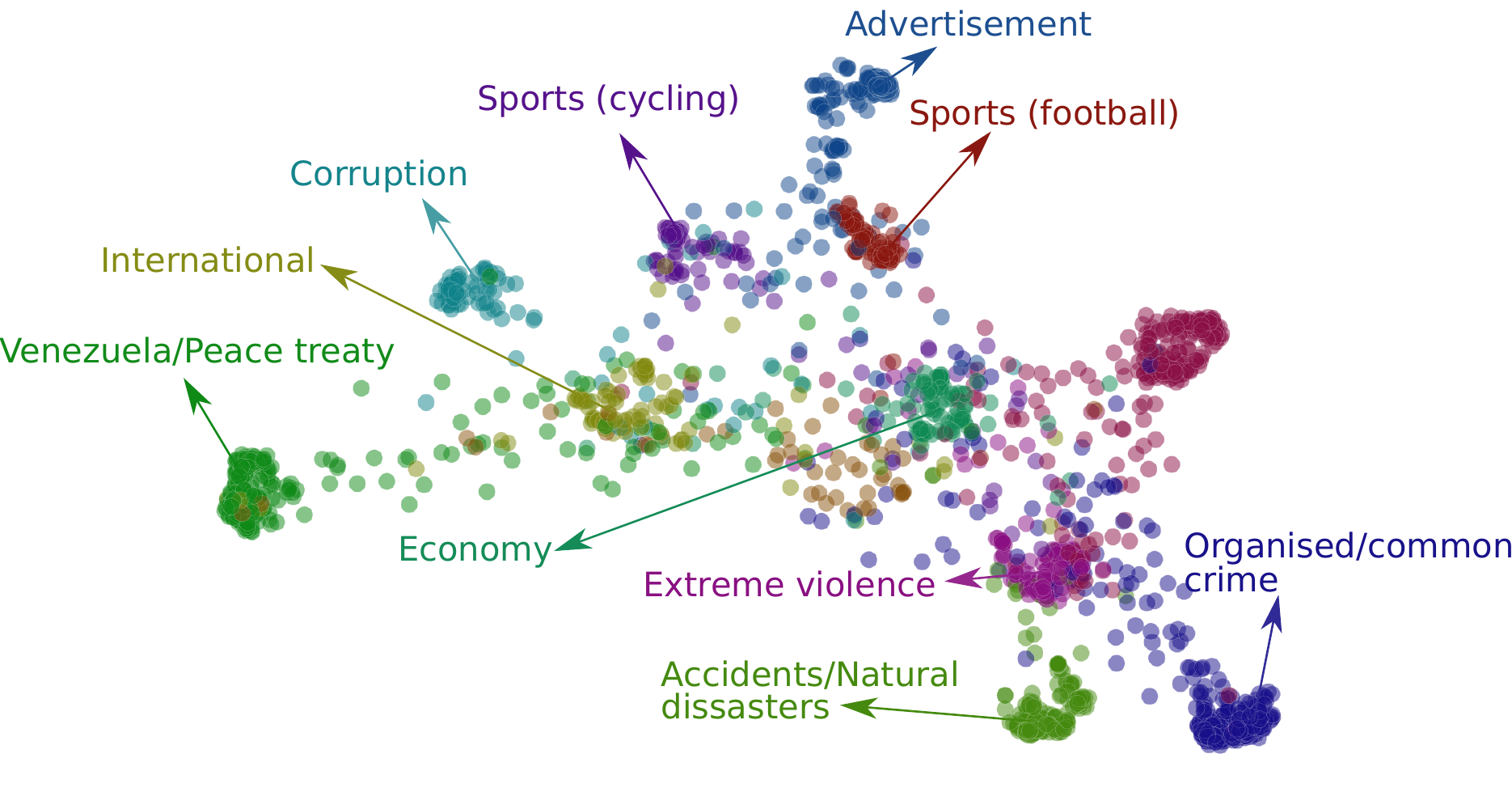}
    \caption{Visualisation of topics discovered with LDA.}
    \label{fig:lda}
\end{figure}

After performing an LDA analysis on the news tweets, each news tweet was assigned a 12 component probability vector. We built the visualisation of such tweets by selecting the tweets that better represented each latent topic. To do this, we imposed a probability threshold on the LDA probability vectors. Tweets whose maximum LDA probability for some topic is above this threshold are called representative tweets for the corresponding latent topic. For a threshold of 0.8, Fig~\ref{fig:lda} shows an annotated visualisation of the most representative tweets. A couple of clusters are not annotated since their topic is not unique or simply not clear.

\begin{figure}[!ht]
    \centering
    \includegraphics[width=\textwidth]{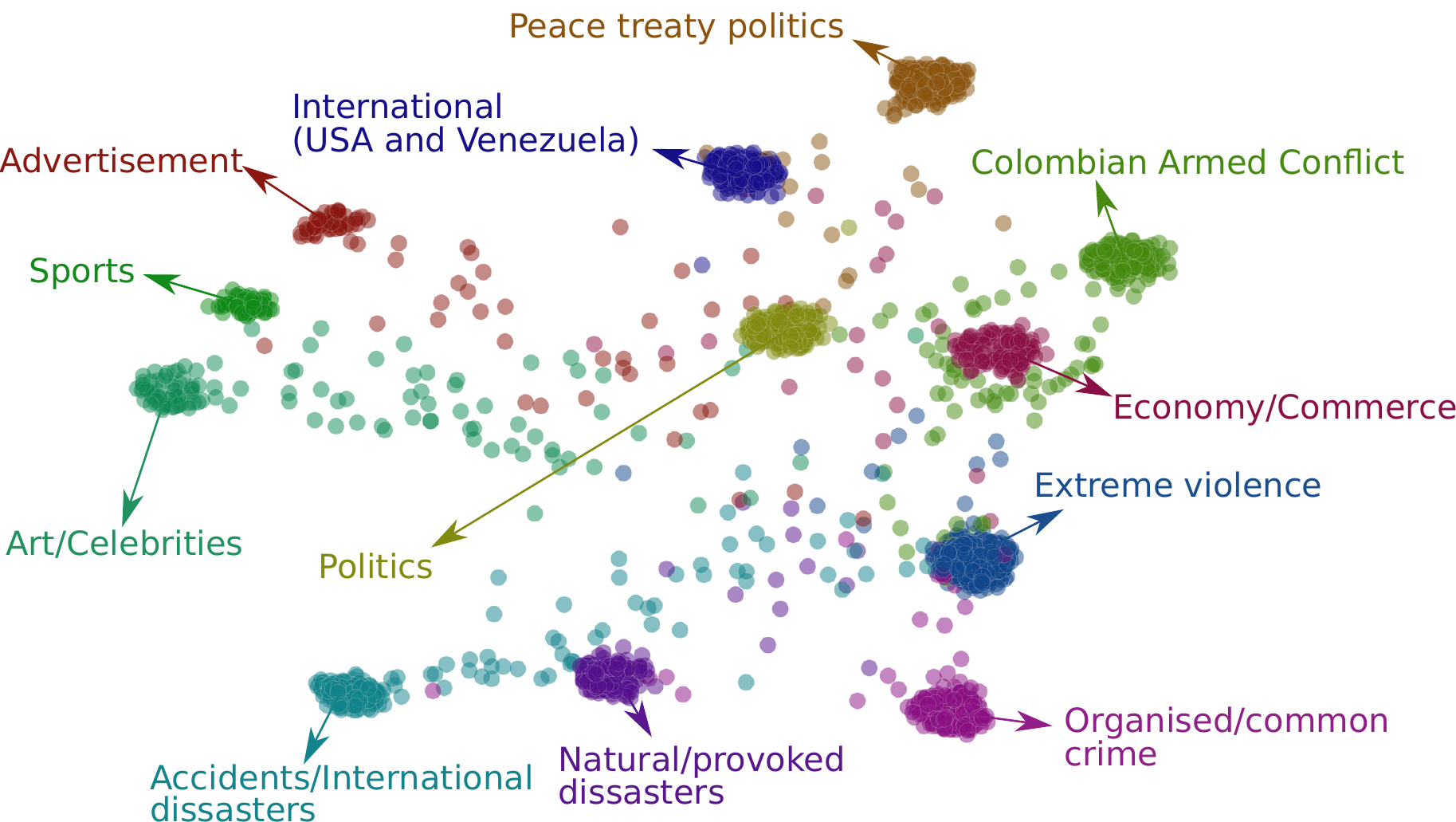}
    \caption{Visualisation of topics discovered with the combination of unsupervised FastText and K-means clustering.}
    \label{fig:kmeans}
\end{figure}

Concerning the FastText and K-means combination, a different selection mechanism for representative tweets was taken into account. K-means groups news tweets geometrically, building 12 vectors called the cluster centroids. We define the most representative tweets of each cluster as the set of tweets grouped in a cluster nearest to its corresponding centroid. Therefore, after defining a maximum distance threshold, we can visualise the clusters and their corresponding representative tweets. Note that this threshold does not have a probabilistic interpretation, and is in general different for different datasets, depending on the mean distance between the data points. In our case, the visualisation is shown in Fig.~\ref{fig:kmeans}. Topics were easier to identify, and the visualisation shows clear clusters well-separated from the others. In terms of visualisation, the FastText + K-means technique is superior for our case study.

\begin{figure}[!ht]%
    \centering
    \subfloat[LDA]{{\includegraphics[width=0.49\textwidth]{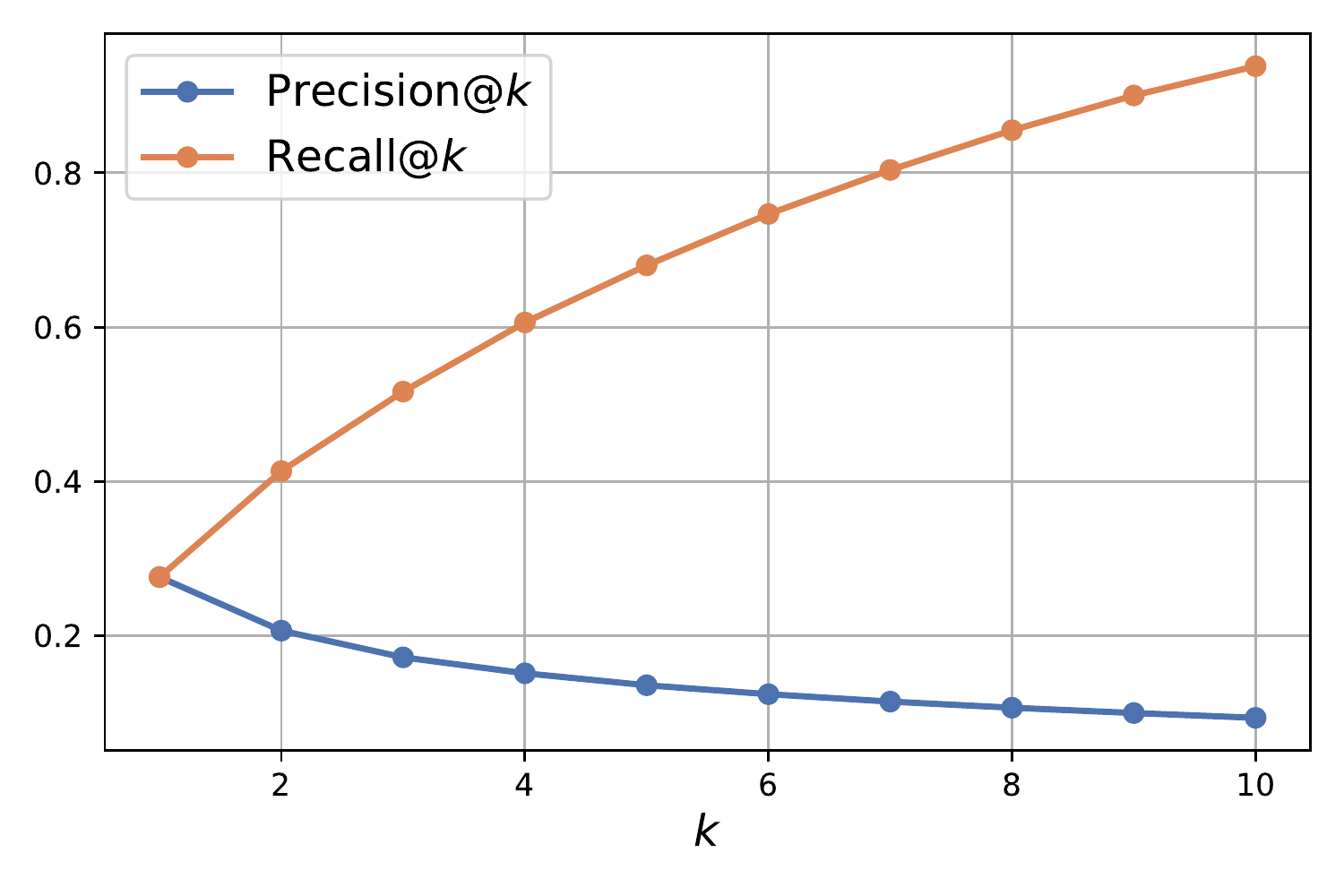} }}%
    \subfloat[FastText + K-means]{{\includegraphics[width=0.49\textwidth]{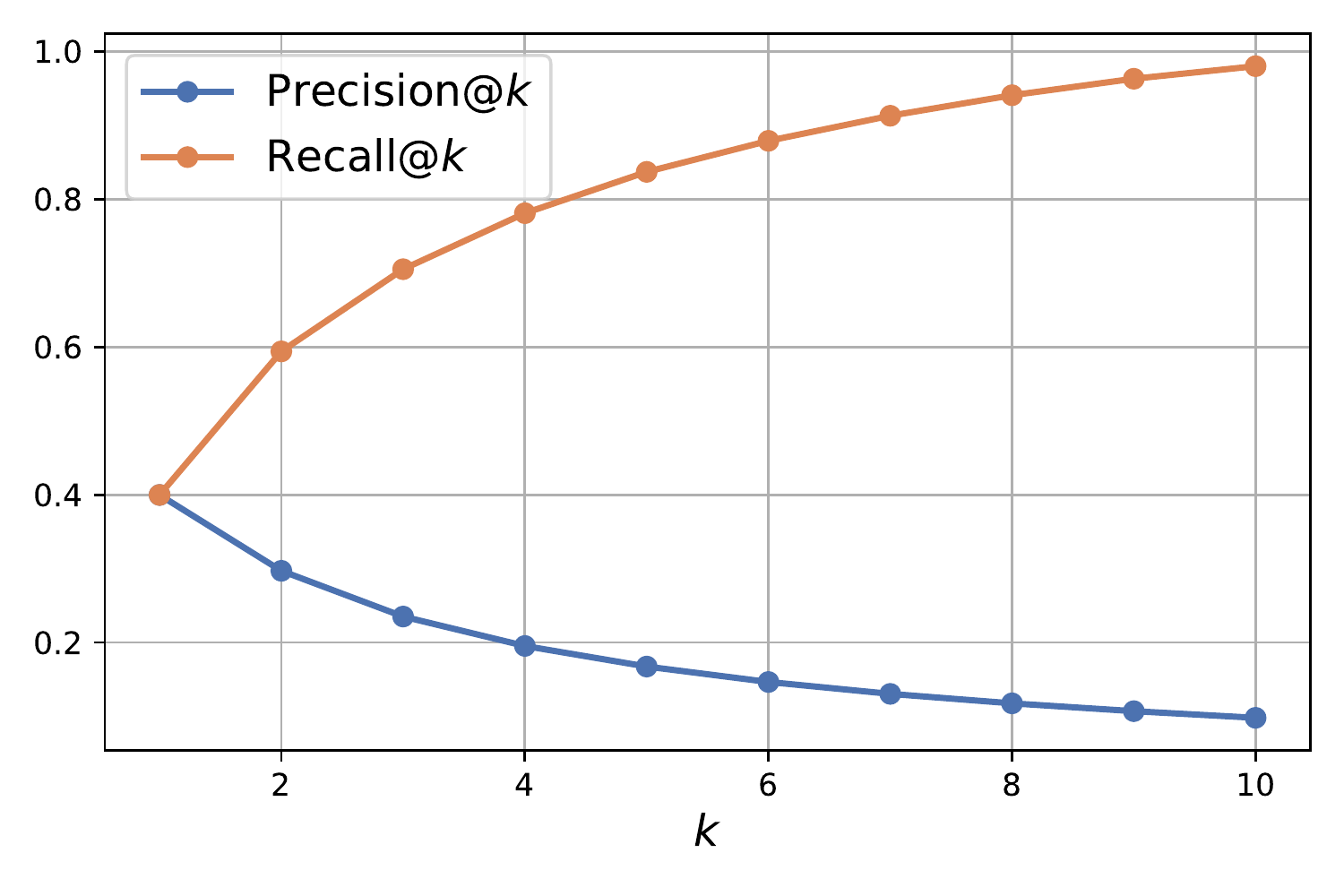} }}%
    \caption{Precision and Recall at $k=1,\ldots,10$ for the LDA and FastText + K-means topic modelling algorithms.}%
    \label{fig:prk}%
\end{figure}

Now, we focus on the second stage of our research, where we predict the topic of a news tweet from a comment to that tweet. The precision and recall at $k=1,\ldots,10$ are shown in Fig.~\ref{fig:prk}. From the results at $k=1$ it is clear that for both LDA and FastText + K-means, the precision and recall are well-above the expected result of a random classifier. Again, FastText + K-means performs better. Of course, as $k$ gets larger, recall increases and precision decreases.

\begin{figure}[!ht]
    \centering
    \includegraphics[width=\textwidth]{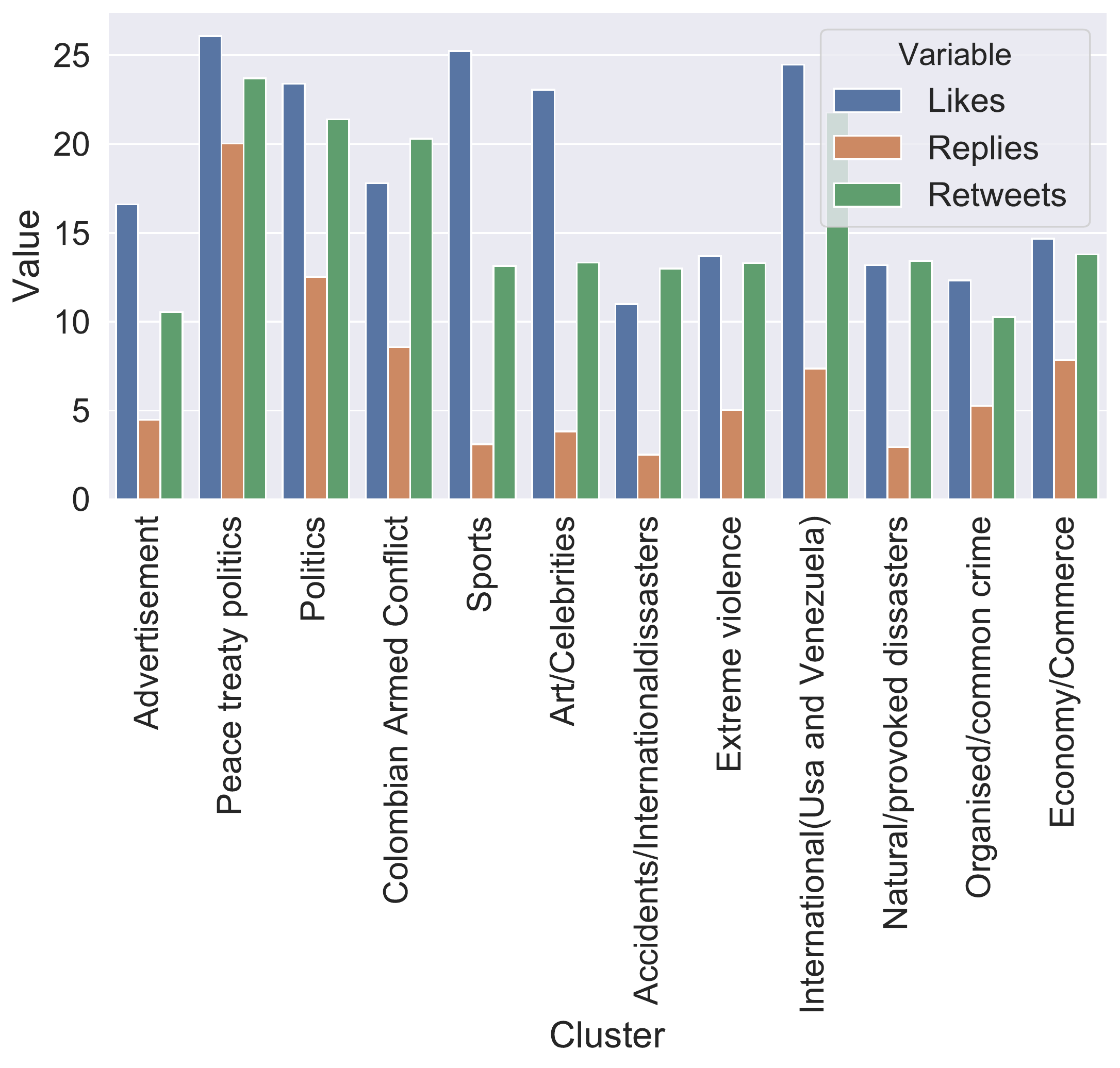}
    \caption{Bar plot of number of likes, replies and retweets for the set of all news tweets belonging to each latent topic discovered by the FastText + K-means method.}
    \label{fig:barplot}
\end{figure}

Remarkably, having a 40\% precision and recall at 1 means that there is a difference between how people respond to different topics. This result is quite good considering that topics were discovered with unsupervised learning, and that documents for classification are single tweets. This difference in response can be further examined with a histogram of the impact caused in the public by each latent topic. We do so only considering the FastText + K-means method, as it has consistently shown to yield better results. Fig. \ref{fig:barplot} shows that the topic identified as peace treaty politics is by far the more engaging one, having the most number of likes, retweets and replies per news tweet related to that topic. Therefore, our analysis confirms that the peace treaty has been the phenomenon with the largest impact on Colombian society.

\section{Conclusions and Future Work}\label{conclusions}

We presented two methods to automatically detect latent topics in news tweets and assess how significant were the differences between the response from the public to tweets from different topics. The first method consisted on discovering latent topics using the widely used LDA model. The second method consisted on generating embedded vectors with FastText and performing K-means clustering on those vectors. We visualised the most representative tweets for each latent topic using both methods and showed that the FastText + K-means method was superior both in the visualisation and in the interpretability of the topics.

Also, using the comments to the news tweets, we trained a supervised FastText method to predict the topic of a news tweet from a comment of that tweet. Again, the better results were obtained with the FastText + K-means method, yielding 40\% precision and recall at 1 in a 12-class classification problem.

The examination of the impact of each topic on the news Twitter account followers revealed that the topic identified as peace treaty politics was the most relevant topic, having the most number of likes, retweets and replies per news tweet, compared to the other topics.

We intend in the future to perform a sentiment analysis on the followers response in order to measure controversiality of topics as well as study which topics evoke positive/negative sentiment in the public.

%
%
%
\bibliographystyle{splncs04}
\bibliography{refs}
\end{document}